\documentclass[aps,superscriptaddress,showpacs,nofootinbib,eqsecnum]{revtex4-1}

\usepackage{amsmath}
\usepackage{mathrsfs}
\usepackage{hyperref}

\usepackage{epsfig}  \input{epsf}

\def\be{\begin{equation}} \def\ee{\end{equation}} \def\bea{\begin{eqnarray}}
\def\eea{\end{eqnarray}}

\def\beq{\begin{equation}}
\def\eeq{\end{equation}}
\def\beqa{\begin{eqnarray}}
\def\eeqa{\end{eqnarray}}

\def\MeV{\rm MeV}

\newcommand{\mev}{{\rm \, MeV}}
\newcommand{\gev}{{\rm \, GeV}}

\newcommand{\avg}[1]{\left\langle#1\right\rangle}

\newcommand{\rpp}{R_{p,\bar{p}}}

\DeclareMathOperator{\Tr}{Tr}

\begin{document}

\title{Evaluation of particle--anti-particle scaled correlation within effective models}

\author{Andr\'{e} F. Garcia} \email{andregarcia@ift.unesp.br}
\affiliation{Instituto de F\'{\i}sica Te\'{o}rica, Universidade Estadual Paulista, 01140-070 S\~ao Paulo, SP, Brazil}

\author{Volker Koch} \email{vkoch@lbl.gov}
\affiliation{Lawrence Berkeley National Laboratory, Berkeley, California 94720, USA}

\author{Marcus B. Pinto} \email{marcus.benghi@ufsc.br}
\affiliation{Departamento de F\'{\i}sica, Universidade Federal de Santa
  Catarina, 88040-900 Florian\'{o}polis, Santa Catarina, Brazil}
\affiliation{Department of Physics, University of Colorado Boulder, Boulder, CO, USA}

\date{\today}

\begin{abstract}

Correlations and fluctuations of physical quantities are known to play an important role in phase transitions and critical phenomena.  In recent years some experimental attempts were made in the scope of the Beam Energy Scan program to locate a possible critical point in the QCD phase diagram. In this work we use the Nambu--Jona-Lasinio model to investigate the off-diagonal quark susceptibility, which is related to the quark--anti-quark scaled correlation at the mean field level. We show that this correlation has a significant peak near the critical point and, therefore, may be a useful quantity to measure in experiment. We further study the effects of a repulsive vector coupling, which reduces the strength of the scaled correlation near the critical point. 

\end{abstract}

\maketitle

\section{Introduction}

The exploration of the QCD phase diagram at nonzero temperature ($T$) and baryonic chemical potential ($\mu_B$) is a matter of great interest and activity in both theoretical and experimental physics. 
It is well known from lattice QCD (LQCD) that, at vanishing baryonic densities, strong interacting matter undergoes an analytic crossover
from the hadronic phase to the quark-gluon plasma \cite{aoki:2006we} at a chiral pseudocritical temperature of $\approx 155\, \rm
MeV$ \cite{aoki:2009sc,Borsanyi:2010bp,bazavov:2011nk}. However, at finite densities the
picture is less clear since there is no reliable information
available from LQCD because of the sign problem \cite{deForcrand:2010ys,Cheng:2007jq}. However,
simulations considering heavy quark masses instead of the physical ones suggest that a first
order phase transition may take place at low temperatures and high
baryonic densities \cite{Ejiri:2008xt}. 
In addition, some effective models of the strong interaction such as the Nambu--Jona-Lasinio model \cite{Nambu:1961tp, Nambu:1961fr, Buballa:2003qv, Klevansky:1992qe, Kunihiro:1987bb}, the linear sigma model \cite{Bowman:2008kc,Ferroni:2010ct} and Polyakov loop models \cite{Fukushima:2017csk, Fukushima:2003fw, Schaefer:2007pw, Ratti:2005jh, Roessner:2006xn, Ratti:2006wg} also predict a first order phase transition at low $T$ and high $\mu_B$. If this conjecture is true, there must be a critical end point (CP) in the phase diagram where the first order phase transition ends.

Many attempts to locate the CP and the first order phase transition region have been made  in recent years, theoretically as well as experimentally. In order to address this issue experimentally, one must be able to scan the QCD phase diagram over a wide region of baryon-number chemical potential. This may be achieved in heavy ion collision experiments by varying the beam energy, $\sqrt{s_{NN}}$, as it is well known that a decrease in the collision energy results in an increase of the baryonic chemical potential (see e.g. \cite{Andronic:2014zha}). A dedicated program to do just that is the \textit{Beam Energy Scan} (BES) program  at the Relativistic Heavy Ion Collider (RHIC). The first phase of the program was completed in $2014$, while the second phase, dedicated to provide data of much improved statistics, is planned to start in 2019 \cite{Odyniec:2015iaa}.
Future experiments such as CBM at FAIR as well as NICA at JINR  are also expected to  probe the
 CP region in the near future. The main goals of the BES program are to search for signals of the first order phase transitions, locate the CP, and determine the conditions in which the QGP signals turn off. In spite of the close connection between the first two goals, the quantities measured to investigate them are quite different. For instance, the value of the directed flow  of net-baryons may indicate a phase transition \cite{Stoecker:2004qu}, as a non-monotonic variation of this quantity is related to a softening of the equation of state \cite{McDonald:2015tza}. In fact, the slope of directed flow of protons and net protons distributions shows a local minimum for energy collisions between $11.5\, \rm GeV$ and $19.6\, \rm GeV$, which may suggest a softer equation of state region \cite{McDonald:2015tza, Stoecker:2004qu, Adamczyk:2014ipa}

The search for the CP, on the other hand, usually focuses on the analysis of fluctuations of various conserved charges, such as baryon number, strangeness and net-charge \cite{Stephanov:1998dy, Stephanov:1999zu, Hatta:2003wn,Jeon:2000wg,Asakawa:2000wh}. Although these fluctuations typically refer to hadronic observables, it is expected that they reflect the thermal properties of the primordial medium. For instance, if the system expands, fluctuations may be frozen in early and thus tell us about the properties of the system prior to its thermal freeze out \cite{Koch:2008ia,Mukherjee:2016kyu}. Special focus has been put on the cumulant ratios of the (net)-baryon distribution, and it has been suggested that non-monotonic behavior of the various  cumulant ratios of the (net)-baryon distribution, such as $\kappa \sigma^2=K_{4}/K_{2}$, with $K_n$ being the n-th order cumulant, could be indicative of a CP \cite{Stephanov:2011pb}. Of course, in practice the fluctuations and correlations of (net)-baryons is difficult to access in heavy ion collision experiments, as they require the detection of neutrons. Therefore, one concentrates on the distribution and correlations of protons and anti-protons as a proxy, which is well justified at least in vicinity of the CP \cite{Hatta:2003wn}. Indeed for energies below $\sqrt{s}\simeq 20 \gev$, the first measurements by the STAR collaboration \cite{Adamczyk:2013dal} show some intriguing beam energy dependence of the kurtosis, $K_{4}/K_{2}$ but improved statistics as well as measurements at even lower energies are needed to draw any firm conclusions.

\begin{figure}[htb]
\begin{center}
\includegraphics[height=3.5cm]{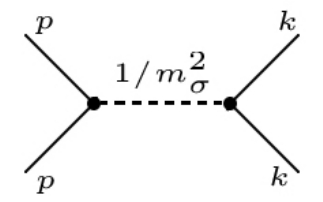}
\caption{Baryon--anti-baryon scattering with the exchange of a sigma mode. }
\label{Fig_Feynman1}
\end{center}
\end{figure}

As the system approaches the critical point,  correlations also play an important role. As pointed out by Stephanov \cite{Stephanov:2004wx}, near the critical point the most singular contribution to the two particle correlator comes from the exchange of the sigma field (Fig. \ref{Fig_Feynman1}). In this case, the correlation length, $\xi$, is governed by the inverse of the sigma meson mass, $\xi \sim 1/m_{\sigma}$, so as the system approaches the critical point $\xi$ experiences a sharp increase as $m_{\sigma} \rightarrow 0$.  For instance, regarding the baryon number, the $1/m_{\sigma}^2$ singularity for the scattering of the two baryons shown in Fig. \ref{Fig_Feynman1} implies the divergence of the baryon number susceptibility. Analogously, given the same conditions, a baryon--anti-baryon scattering implies a similar divergence of the baryon--anti-baryon number correlation. 
Indeed, the correlations between baryons and anti-baryons may turn out to be a more sensitive probe than the cumulants of the (net)-baryon number. In the absence of any interaction, baryon--anti-baryon correlations vanish while the cumulants still retain the finite value of a Poisson distribution, $K_{n}\sim \avg{N}$. Therefore, the correlations between baryons and anti-baryons will provide interesting, complementary, and possibly more sensitive information about the phase structure of QCD. And, as we shall discuss in detail, the presently available STAR data already contain, albeit statistically not very significant,  information about the correlations between protons and anti-protons.  Therefore, it is interesting to calculate, in an effective model, the expected strength of proton--anti-proton correlations close to the (pseudo) critical transition region. This is the main purpose of the present paper.

This paper is organized as follows. In the next section we present a few remarks on the correlation function and the correlation coefficient (or scaled co-variance) that we are seeking to evaluate in the NJL model and explain how to extract the $p-\bar{p}$ correlation from the data published by the STAR Collaboration. In section III we introduce the Nambu--Jona-Lasinio model and evaluate the quark--anti-quark number correlation within the mean field approximation (MFA). In section IV we present and discuss our results.

\section{Particle--anti-particle correlation and the STAR data}
\label{Sec2}

Usually, correlations between particles are measured directly by comparing the pair distribution  with the product of single particle distribution for the particles involved. However, the (integrated) correlation can also be extracted from cumulant measurements. For example, consider the second order net-proton ($\Delta N_{p}=N_{p}-N_{\bar{p}}$) cumulant
\begin{eqnarray}
  K_{2}(\Delta N_{p})=\avg{\delta(N_{p}-N_{\bar{p}})^{2}}=\avg{(\delta N_{p})^{2}}+\avg{(\delta N_{\bar{p}})^{2}} - 2\avg{\delta N_{p} \,\delta N_{\bar{p}}}.
  \label{eq:k2_prot}
\end{eqnarray}
Clearly, the correlation $C_{p,\bar{p}}=\avg{\delta N_{p} \, \delta N_{\bar{p}}}$ between protons and anti-protons is given by the difference between the net proton cumulants and those for protons, $K_n(N_p)$, and anti-protons, $K_n(N_{\bar{p}})$,
\beq
C_{p,\bar{p}}= \frac{1}{2} \left( K_{2}(N_p)+K_{2}(N_{\bar p})-K_{2}(\Delta N_p) \right).
\vspace{0,4 cm}
\label{Eq_C2correlator}
\eeq

During the first phase of the BES program the STAR Collaboration actually measured the net-proton, proton and anti-proton cumulants in \textit{Au-Au} collisions  up to forth order and for a wide range of beam energies \cite{Adamczyk:2013dal}. 

Cumulants are extensive quantities, i.e. they depend on the volume of the system, which is not very well known for a heavy ion collision. To avoid this problem, one commonly considers ratios of cumulants. In our case, we consider the following scaled co-variance, 
\beq
R_{p,\bar{p}}=\frac{C_{p,\bar{p}}}{\sqrt{K_1(N_p) K_1(N_{\bar{p}})}}\, ,
\vspace{0,4 cm}
\label{Eq_R1}
\eeq
\noindent where $K_1(N_{p})=\avg{N_{p}}$ and $K_1(N_{\bar p})=\avg{N_{\bar{p}}}$ are the first order cumulants, i.e. the means of the proton and anti-proton distributions, respectively. From now on we will refer to $R_{p,\bar p}$ simply as the proton--anti-proton scaled correlation or scaled co-variance. 

\begin{figure}[htb]
\begin{center}
\includegraphics[height=6.0cm]{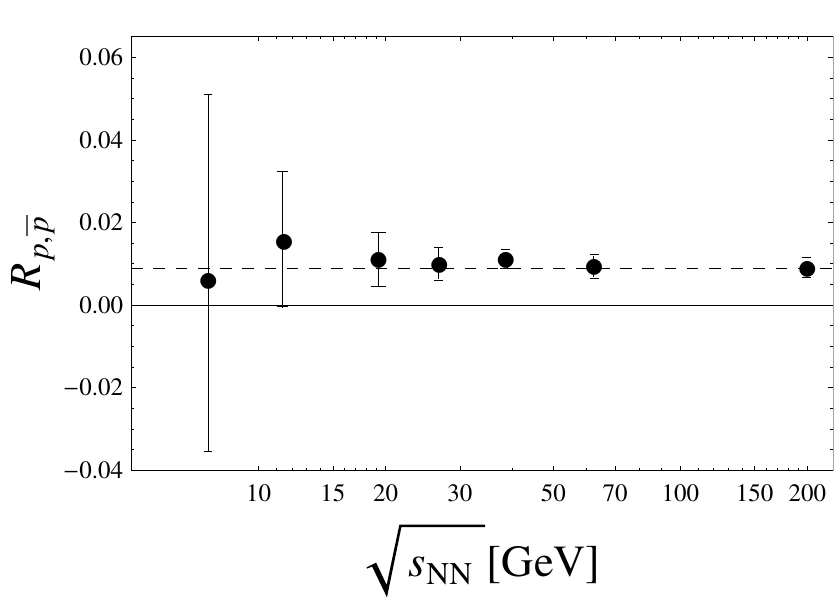}
\caption{Proton--anti-proton scaled correlation, $R_{p, \bar p}$, with error bars at the different values of $\sqrt{s_{NN}}$. The dashed line represents a rough horizontal fit to the data to give an approximate idea about the magnitude of the correlation. The values of $R_{p, \bar p}$ were calculated using data taken from the STAR Collaboration \cite{Adamczyk:2013dal, STARsite}.}
\label{Fig_RppData}
\end{center}
\end{figure}

Using Eqs. (\ref{Eq_R1}) and (\ref{Eq_C2correlator}) we can easily construct the corresponding proton--anti-proton scaled correlation for the STAR data as shown in Table  \ref{Table_3} and Fig. \ref{Fig_RppData}. We see that at $\sqrt{s_{NN}}=7.7\, \rm GeV$ and $\sqrt{s_{NN}}=11.5\, \rm GeV$ the correlations, within errors, are consistent with zero, whereas for the higher energies, the data show a significant deviation from zero. Overall, given the large errors at low energies, we observe that the scaled correlation, $R_{p,\bar{p}}$, is essentially independent of the collision energy, which is indicated by the dashed horizontal line in Fig. \ref{Fig_RppData} which represents a rough fit to the STAR data used for comparison with our model calculations.

\begin{table}[htb]	
\begin{center}
\begin{tabular}{c||c}
$\sqrt{s_{NN}}~ (\textrm{GeV})$ &
$R_{p,\bar{p}}$    \\ \hline
$7.7$~            &  ~$0.00640785 \pm 0.0432968$~    \\ \hline
 $11.5$         &     $0.01583915 \pm 0.016335$    \\ \hline
$19.6$ & $0.0107651 \pm 0.0065406$  \\ \hline
 $27$ & $0.009361 \pm 0.00381434$  \\ \hline
 $39$ & $0.0111666 \pm 0.001944095$  \\ \hline
 $62.4$ & $0.00943055 \pm 0.002464215$  \\ \hline
 $200$ & $0.00894935 \pm 0.001132705$  \\ \hline
\end{tabular}
\end{center}
\caption{The proton--anti-proton scaled correlation (right column), given by Eq. (\ref{Eq_R1}), at different collision energies (left column) evaluated from the STAR data \cite{Adamczyk:2013dal, STARsite}.}
\label{Table_3}
\end{table}

Despite the lack of a clear peak in the data, correlations do play an important role in the vicinity of a critical point. Thus it would be interesting to explore how -- in an effective model -- the above proton--anti-proton correlation would behave in the various regions of the phase diagram and, in particular, in the vicinity of a critical point. To this end let us briefly review how cumulants are commonly calculated in an effective model at finite temperature. 

Effective models are useful to describe strong interacting matter in thermal equilibrium. Such a system may be characterized by its partition function, $Z$, which is a function of the Hamiltonian of the system, $H$, the conserved charges, $Q_i$, and their respective chemical potential, $\mu_i$,

\beq
Z= \Tr \left[ \exp \left( -\frac{H-\sum_i \mu_i Q_i}{T} \right) \right] \, .
\eeq

Statistical quantities, such as the mean and the (co)-variances, are then obtained as derivatives of the partition function with  respect to the appropriate chemical potential(s), 

\beq
\langle Q_i \rangle = T \frac{\partial}{\partial \mu_i} \log (Z)\, ,
\eeq

\beq
\langle \delta Q_i \delta Q_j \rangle = T^2 \frac{\partial^2}{\partial \mu_i \partial \mu_j} \log (Z)=VT \chi_{i,j} \, ,
\label{eq:charge_cumulant}
\eeq
with the susceptibilities $\chi_{i,j}$ given by
\beq
\chi_{i,j}=\frac{T}{V} \frac{\partial^2}{\partial \mu_i \partial \mu_j} \log (Z)\, .
\eeq

\noindent The diagonal susceptibilities, $\chi_{i,i}$, are a measure for the fluctuations of the system, whereas the off-diagonal susceptibilities, $\chi_{i,j}$, with $i\neq j$, characterize the correlations between conserved charges $Q_i$ and $Q_j$. Susceptibilities are related to integrals of equal time correlation functions of the appropriate charge-densities. In this work we will concentrate on  second order susceptibilities. If we consider the density fluctuation $\delta \rho_i(x)=\rho_i(x)- \bar{\rho}_i$, with $\bar{\rho}_i$ being the spatially averaged density of the charge $Q_i$, then the relation between the second order susceptibility and the density-density correlation function is given by

\beq
\chi_{i,j}=\frac{1}{VT}\int d^3xd^3y \langle \delta \rho_i(x) \delta \rho_j(y) \rangle = \frac{1}{T} \bar{\rho}_i \delta_{i,j} + \frac{1}{T}\int d^3r C_{i,j}(r)\, ,
\eeq

\noindent where $C_{i,j}(r)$ are the correlation functions,

\beq
C_{i,j}(r)=\langle \delta \rho_i(r) \delta \rho_j(0) \rangle - \bar{\rho}_i \delta_{i,j} \delta (r) \sim \frac{\exp [-r/\xi_{i,j}]}{r}\, .
\eeq

\noindent The correlation length, $\xi_{i,j}$, provides a measure for the strength and type of the correlation. 
In this work, we are concerned with the correlation between particle and anti-particle numbers and therefore, we will need the correlation function which depends on the density functions of particles and anti-particles. In principle this is a problem, since in full theories one is only able to evaluate susceptibilities of conserved quantities, such as the net baryon or net electric charge. Thus quantities that are not related to net (conserved) charges are not readily accessible  in thermal field theory. The proton--anti-proton number susceptibility, for instance, is such a case. However, in the mean field approximation the particle and anti-particle distribution functions are independent, so that the (off-diagonal) susceptibility of particles and anti-particles numbers may be easily evaluated. This means that within the mean field approximation one is able to calculate the effects of the sigma exchange between baryons and anti-baryons. Since the argument based on the simple sigma-exchange discussed above is rather generic, we believe that such a calculation contains the relevant physics and thus provides an important estimate for the strength of correlations to be expected.

\begin{figure}[htb]
\begin{center}
\includegraphics[height=3.0cm]{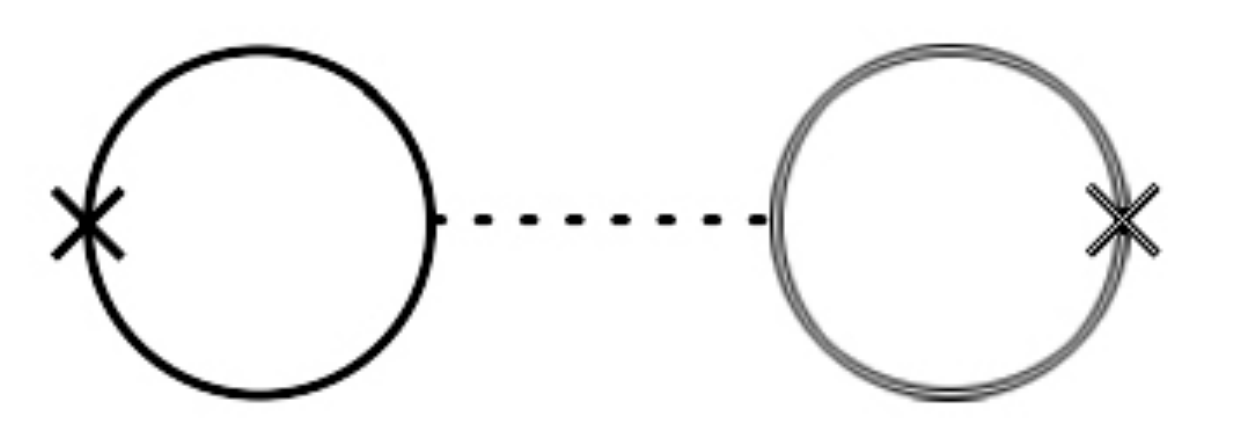}
\caption{Contribution to the particle--anti-particle second order susceptibility.}
\label{Fig_Feynman2}
\end{center}
\end{figure}

Finally, as already pointed out, in statistical equilibrium it is useful to work with the scaled (co)-variance, or correlation coefficient, which can be expressed as a ratio of susceptibilities \cite{Koch:2005vg}. Since we are interested in the particle--anti-particle scaled correlation, we may express the correlation coefficient as a ratio of the particle--anti-particle number susceptibility and their respective densities,

\beq
R_{i,j}=\frac{T \chi_{i,j}}{ \sqrt{\rho_i \rho_j}}\, ,
\label{Eq_R210}
\eeq
\noindent where $\rho_i = \langle Q_i \rangle/V$ is the particle number density ($i \neq j$). The susceptibility $\chi_{i,j}$ may be represented by the diagram in Fig. \ref{Fig_Feynman2}, where the left loop represents a particle while the right loop represents an anti-particle.

Thus, within the mean field approximation, it is possible to calculate the scaled particle--anti-particle fluctuations and correlations. This then will allow at least a qualitative exploration of these quantities in the 
$T-\mu$ phase diagram and, in particular, in the vicinity of the critical point. In this work, we will employ the well known Nambu--Jona-Lasinio (NJL) model \cite{Nambu:1961tp, Nambu:1961fr} for these calculations, which we will discuss in the next section.



\section{Evaluation of the Scaled Correlation}
\label{method}

Let us now evaluate the scaled correlation given by Eq. (\ref{Eq_R210}) using the  NJL model. As discussed in the previous section, Eq.~\eqref{eq:charge_cumulant}, the second order cumulant of the baryon number distribution is related to the baryon number susceptibility by 

\begin{align}
  K_2=\frac{1}{2T^2}\chi_{B, B}\, .
\end{align}
Here, we are interested in the baryon--anti-baryon number correlation, which is related to the off-diagonal baryon--anti-baryon susceptibility in the same way \cite{Koch:2008ia},
\begin{align}
  C_{B, \bar B}=\frac{1}{2T^2}\chi_{B, \bar B}\, .
  \label{Eq_C2Xqq}
\end{align}

As already pointed out, in the mean field approximation the baryon--anti-baryon susceptibility is well defined and we will now discuss how it is obtained in the NJL model
 whose simplest version  is described by a Lagrangian density for fermionic fields given by~\cite{Nambu:1961tp, Nambu:1961fr, Buballa:2003qv, Klevansky:1992qe}

\begin{equation}
\mathcal{L}_{\rm NJL}={\bar \psi}\left( i{\partial \hbox{$\!\!\!/$}}-m\right) \psi
+G\left[ ({\bar \psi}\psi)^{2}-({\bar{\psi}} \gamma _{5}{\vec{\tau}}\psi
  )^{2}\right] ,
\label{njl2}
\end{equation}
where $\psi$ (a sum over flavors and color degrees of freedom is implicit)
represents the flavor iso-doublet ($u,d$ quarks) and $N_{c}$-plet quark
fields, $\vec{\tau}$ are isospin Pauli matrices, and $m$ is the current quark mass (which we assume to be the same for  both \textit{up} and \textit{down} quarks). 
The Lagrangian density
(\ref{njl2}) is invariant under (global) $U(2)_{\rm f}\times SU(N_{c})$ and,
when $m=0$, the theory is also invariant under chiral $SU(2)_{L}\times
SU(2)_{R}$. 
Within the NJL model a sharp cut off ($\Lambda$) is generally used as an ultra violet regulator and since the
model is non-renormalizable,  $\Lambda$, together with the coupling constant $G$ and the current quark mass $m$, are  parameters of the model which need to be fixed.  This is done by requiring that the phenomenological values
the pion mass ($m_{\pi}$),  the pion decay constant
$(f_{\pi})$, and the quark condensate ($ \langle {\bar \psi} \psi \rangle$) be reproduced. 
Here, we choose the set
 $\Lambda=590\,\rm MeV$ and $G\Lambda^2=2.435$ with $m=6\,\rm MeV$
in order to reproduce $f_\pi=92.6\,\MeV$, $m_\pi= 140.2\,\MeV$, and $ \langle {\bar \psi} \psi \rangle^{1/3}=-241.5 \, \MeV$ \cite {Frank:2003ve}.




Given the above Lagrangian, at finite temperature and chemical potential the mean field thermodynamical potential may be written as (see Ref. \cite {Ferroni:2010ct, Kneur:2010yv} for results beyond MFA)


\beq
\Omega(T,\mu_q, \mu_{\bar{q}}, M)=\frac{(M-m)^2}{4G}-2N_cN_f\int \frac{d^3p}{(2\pi)^3}\left \{E_p+T \ln [1+e^{-(E_p-\mu_q)/T}]+T \ln [1+e^{-(E_p-\mu_{\bar{q}})/T}] \right \}\, ,
\vspace{0,4 cm}
\label{Omega2}
\eeq

\noindent where $M$ is the effective quark mass. The first term in the above integral is the vacuum contribution, which must be regularized by $\Lambda$, while the second and third terms are the particle (quark) and anti-particle (anti-quark) contributions with chemical potential $\mu_q$ and $\mu_{\bar q}$, respectively. In order to make the system thermodynamically consistent we must set $\mu_q=\mu$ and $\mu_{\bar q}=-\mu$ at every numerical evaluation, with $\mu$ being the usual chemical potential for net quark number. 

The pressure is the negative of the thermodynamic potential evaluated at the solution $M^*$ of the gap equation, $p(T,\mu_q,\mu_{\bar{q}})=-\Omega(T,\mu_q,\mu_{\bar{q}},M^*)$. 
The off-diagonal quark--anti-quark number susceptibility, $\chi_{q,\bar{q}}$, is given by the second order derivative of $\Omega$ with respect to $\mu_q$ and $\mu_{\bar q}$,

\beq
\chi_{q,\bar{q}}=\frac{1}{T \hat V} \langle \delta N_q \delta N_{\bar{q}} \rangle= -\frac{d^2 \Omega}{d \mu_q d \mu_{\bar{q}}}\, .
\label{Chi2}
\eeq

Equations (\ref{Omega2}) and (\ref{Chi2}) allow us to evaluate the quark--anti-quark scaled correlation. Working out the derivatives in Eq. (\ref{Chi2}), including the implicit ones (for details see Appendix \ref{AppXq}), we find that the off-diagonal quark number susceptibility may be written as



\beq
\chi_{q, \bar q}=\frac{1}{\Omega^{\prime \prime}_M}\frac{\partial \rho_q}{\partial M}\frac{\partial \rho_{\bar q}}{\partial M}\, ,
\label{Eq_Chi3}
\eeq

\noindent where $\rho_q$ and $\rho_{\bar q}$ are quark and anti-quark number densities,

\beq
\rho_q=-\frac{\partial \Omega}{\partial \mu_q} \, , \,\,\,\,\,\,\,\,\,\,\,\,\,\,\,\,\,\,\,\,\,\, \rho_{\bar q}=-\frac{\partial \Omega}{\partial \mu_{\bar q}}\, ,
\label{Eq_Densities}
\eeq

\noindent and $\Omega^{\prime \prime}_M$ is defined as

\beq
\Omega^{\prime \prime}_M=\frac{\partial^2 \Omega}{\partial M^2} \, .
\eeq

Given the quark--anti-quark susceptibility, Eq.~(\ref{Eq_Chi3}), we next need to relate it to the proton--anti-proton susceptibility,  $\chi_{p, \bar{p}}$. First, we note that the baryon chemical potential is three times that of the quark chemical potential, $\mu_B=3\mu_q$. Therefore, the baryon--anti-baryon number susceptibility is related to the quark--anti-quark number susceptibility by a factor of $1/9$. Also, if we take into account isospin symmetry, we find that the relation between proton--anti-proton and quark--anti-quark susceptiblities is given by







\beq
\chi_{p, \bar{p}}= \frac{1}{36}\chi_{q, \bar{q}}\, .
\label{Eq_Chi36}
\eeq

Using the same arguments we can see that the quark density is related to the proton density by a factor of $1/6$, 







\beq
\rho_p=\frac{1}{6}\rho_q\, .
\label{Eq_RhoqRhop}
\eeq

\noindent Replacing equations (\ref{Eq_RhoqRhop}) and (\ref{Eq_Chi36}) into equation (\ref{Eq_R1}) we obtain the proton--anti-proton scaled correlation written in terms of quark and anti-quark quantities that can be easily evaluated in the NJL model,

\beq
R_{p, \bar{p}}=\frac{1}{12}\frac{T\chi_{q, \bar{q}}}{\sqrt{\rho_q \rho_{\bar{q}}}}\, .
\vspace{0,4 cm}
\label{Eq_R2}
\eeq

Equation (\ref{Eq_R2}) allow us to numerically evaluate the proton--anti-proton scaled correlation as a function of $T$ and $\mu$ within an effective model. Meanwhile the data published by STAR for the first and second order cumulants of proton and anti-proton distributions \cite{Adamczyk:2013dal, STARsite} give us access to this type of correlation for some energy collision values (see Table \ref{Table_3}). This way we can compare the experimental data to  results evaluated from effective models.

\section{Numerical Results}

\begin{figure}[htb]
\begin{center}
\includegraphics[height=6.0cm]{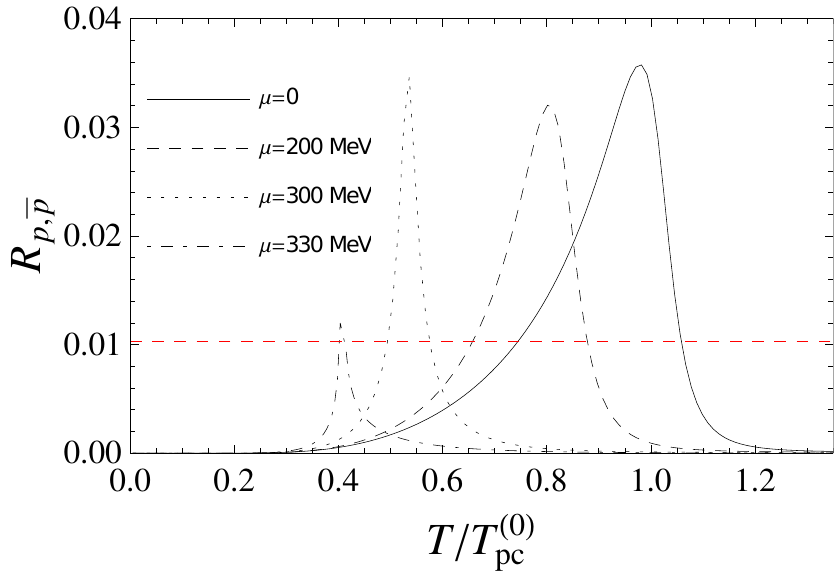}

\caption{Proton--anti-proton scaled correlation evaluated within the NJL model as a function of the temperature normalized by the pseudo-critical temperature at $\mu=0$. The solid line shows the  $\mu=0$ case while the black dashed and dotted lines show the $\mu=200\, \rm  MeV$  and $\mu=300\, \rm MeV$ cases, respectively. The dot-dashed line shows the case for $\mu=330\, \rm MeV$, which is greater than the critical chemical potential, $\mu_c \simeq 328\, \rm MeV$. The red dashed line is a linear fit of the STAR data (see Fig.~\ref{Fig_RppData}).
}
\label{Fig_Gv0_RvsT}
\end{center}
\end{figure}

\begin{figure}[htb]
\begin{center}
\includegraphics[height=5.5cm]{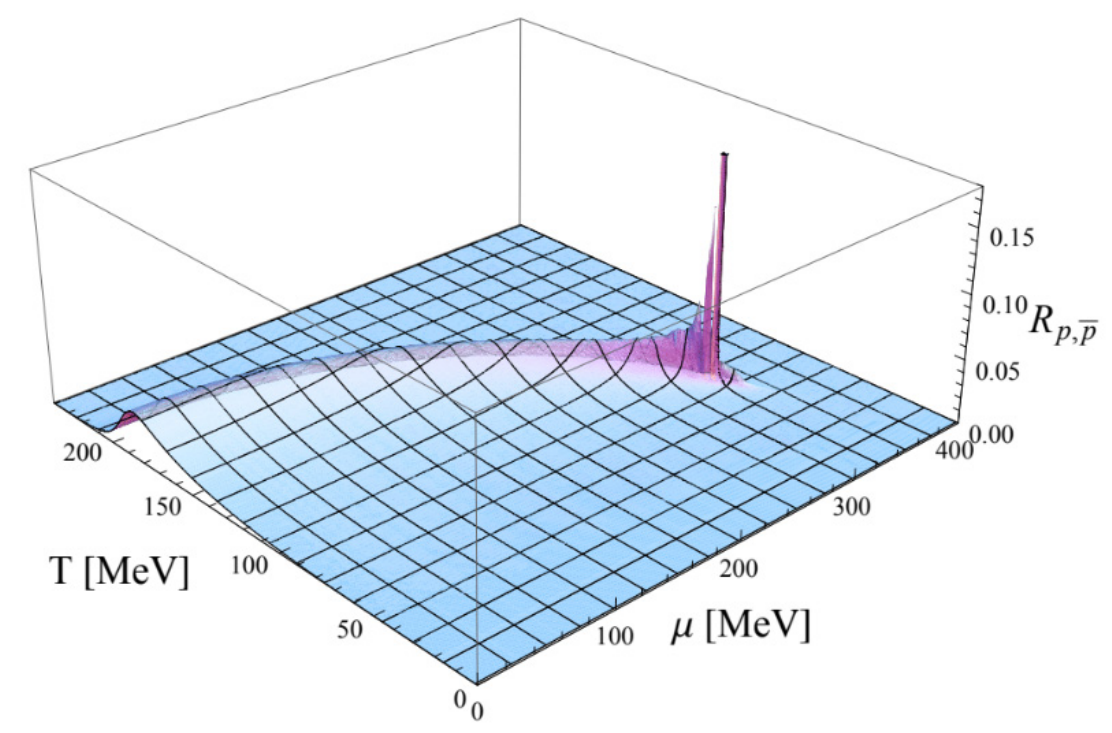}
\caption{Proton--anti-proton scaled correlation in the $T-\mu$ plane.}
\label{Fig_Gv0_R3D}
\end{center}
\end{figure}

\begin{figure}[htb]
\begin{center}
\includegraphics[height=5.5cm]{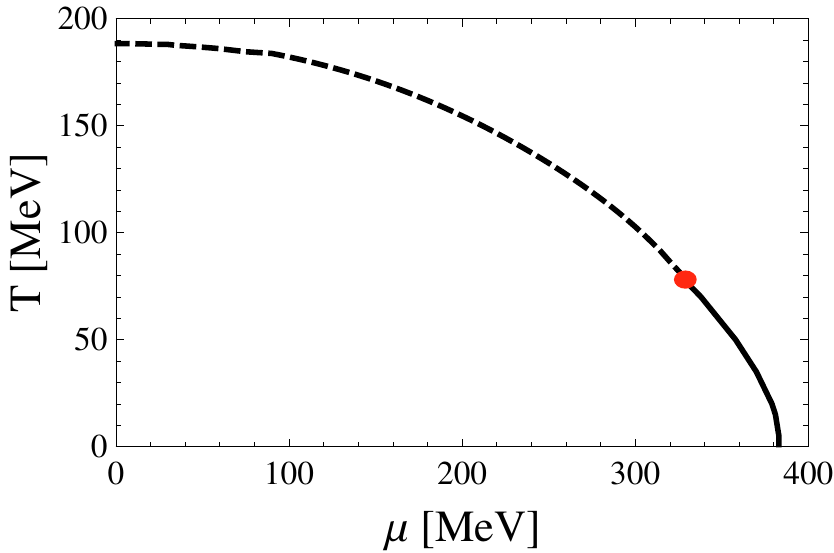}
\includegraphics[height=5.5cm]{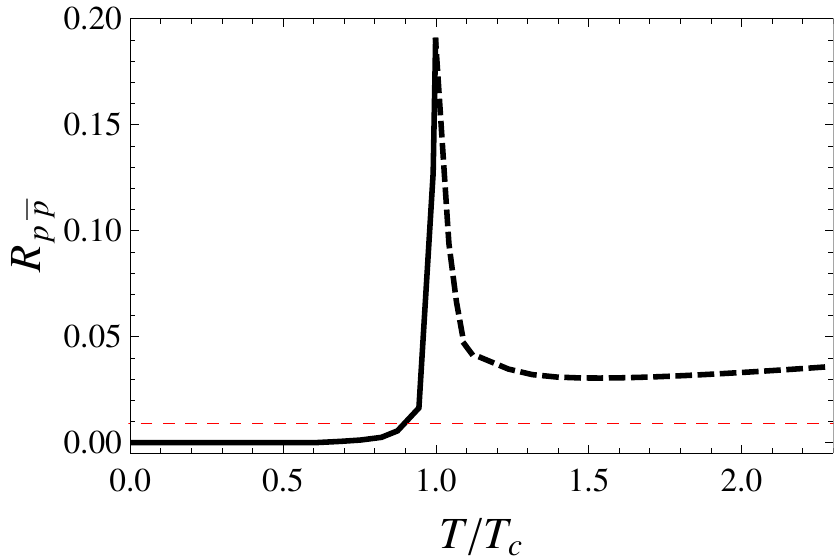}
\includegraphics[height=5.5cm]{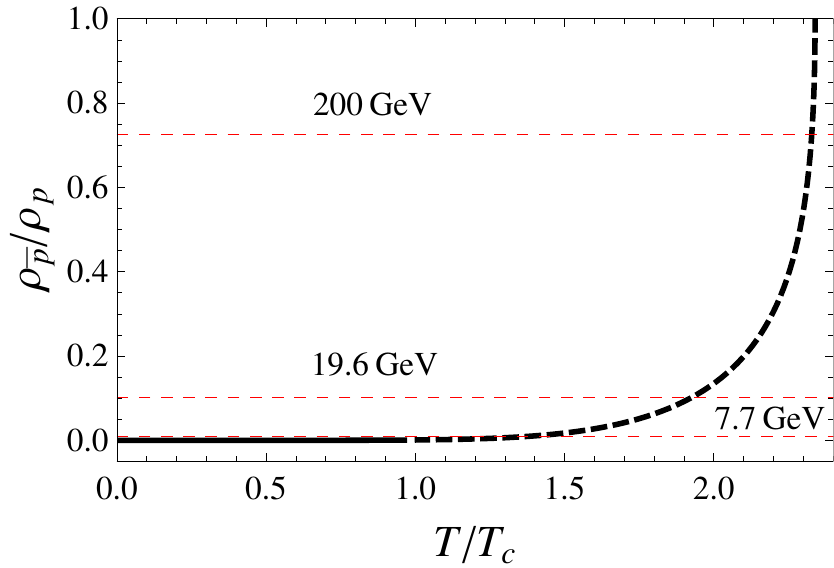}
\caption{Top left: The model phase diagram. The dashed  line corresponds to the cross-over pseudo-critical temperature, and the full line represents the first order co-existence. Top right: Proton--anti-proton scaled correlation along the phase transition line as a function of $T/T_c$, where $T_c$ is the critical temperature. The red dashed line represents our rough fit to the STAR data, Fig.~\ref{Fig_RppData}. Bottom: Ratio of anti-protons over protons along the (pseudo)-critical line as given in our model. The red dashed lines represent the ratio evaluated from STAR data for the beam energies, $\sqrt{s_{NN}}$, as indicated in the plot.}
\label{Fig_Gv0_RPD}
\end{center}
\end{figure}

Using equations (\ref{Eq_Chi3}), (\ref{Eq_Densities}) and (\ref{Eq_R2}) we can evaluate the proton--anti-proton scaled correlation in terms of quarks and anti-quarks suscepitiblities and densities within the NJL model. The results for different values of the chemical potential are shown in  Fig.~\ref{Fig_Gv0_RvsT}, which displays the scaled correlation as a function of the temperature. The full line represents the $\mu=0$ case, where we see that $R_{p, \bar{p}}$ is close to zero at low $T$. It grows rapidly as $T$ increases and then peaks at the crossover pseudo-critical temperature\footnote{We define the crossover pseudo-critical temperature as the temperature that maximizes the value of $-\partial M/\partial T$ at a fixed chemical potential. On the other hand, the use of susceptibilities may be usefull to define a band of pseudo-critical temperatures \cite{Du:2013oza}.}, $T_{pc}^{(0)}=188\, \rm MeV$, before  dropping to values close to zero at higher temperatures. At finite baryonic densities a  similar behavior is also observed and, for a given chemical potential value,  the  scaled correlation peaks at the corresponding pseudo-critical temperature, $T_{pc}^{(\mu)}$. We also see a slight decrease of the peak value for moderate chemical potentials ($\mu \lesssim 250\, \rm MeV$) . However, as we approach the critical point the peak sharpens and the maximum increases before it drops rapidly for $\mu >\mu_{c}$. The overall behavior is illustrated in Fig.~\ref{Fig_Gv0_R3D}, where we show the scaled correlation for the entire  $T-\mu$ plane. We see that the scaled correlation between protons and anti-protons exhibits a well defined maximum along the pseudo-critical line, which turns into singularity at the critical point.\footnote{The fact that $\rpp$ appears to be finite in Fig.~\ref{Fig_Gv0_R3D} is simply due to the finite resolution in $T$ and $\mu$ in our numerical calculation.} We also find  that the value of the maximum along the pseudo-critical line changes only very mildly. This mild dependence is seen even better in the top-right panel of Fig.~\ref{Fig_Gv0_RPD} where we plot the scaled correlation along the (pseudo)-critical line, which is depicted in the upper left panel of the same figure. For comparison we also show, as the red dashed line, the rough fit to the STAR data from Fig.~\ref{Fig_RppData}. The position of the critical point is indicated by the red dot, noting again that the apparent finite value of $\rpp$ at the critical point is simply due to the finite resolution in our calculation.

The peak of the correlation observed in Figs. \ref{Fig_Gv0_R3D} and \ref{Fig_Gv0_RPD} (right panel) suggests that the proton--anti-proton scaled correlation provides a good signal to identify the model's critical point. Also noteworthy is that $\rpp$ is essentially constant along the pseudo-critical line (top right panel of Fig.~\ref{Fig_Gv0_RPD}).
This is consistent with the behavior seen in the 
STAR data, suggesting that the STAR data probeb a region close to the pseudo-critical line, as one  would expect from the analysis of particle ratios \cite{Andronic:2014zha,Adamczyk:2017iwn}.  The fact that the model predicts a value of $\rpp$ which is considerably larger than that seen in the data should  be of no concern given the simplicity of the present model. It rather suggests that it would be worthwhile to explore $\rpp$ in more complete approximations to QCD such as the Dyson-Schwinger \cite{Fischer:2014mda} or functional renormalization group approach \cite{Herbst:2013ufa}. However, the qualitative trend of a rapid increase of $\rpp$ in the vicinity of the critical point is genuine and rather model independent as already discussed previously. Unfortunately, the presently available STAR data at the lowest beam energies have such large error-bars that no conclusions can be drawn. For completeness we also show, in the bottom panel of Fig.~\ref{Fig_Gv0_RPD}, the ratio of anti-protons over protons along the pseudo-critical line together with the values from the STAR measurement for various energies. Despite the simplicity of the NJL model it is encouraging to see that it explores roughly the same region for this ratio.

Of course, the location of the critical point and the crossover pseudo-critical line is clearly model dependent. In order to explore some of the model dependence, we now investigate the scaled correlation considering a repulsive vector coupling within the NJL model, which is known to weaken the first order phase transition while lowering the critical temperature and increasing the critical chemical potential values \cite{Kitazawa:2003qmg, Fukushima:2008wg}. We will see how the scaled correlation behaves as the location of the critical point  changes when different vector coupling values are considered. As we go to lower temperatures in the phase diagram, the anti-particle contribution to the thermodynamic potential becomes less important and we expect this to reflect in the proton--anti-proton scaled correlation.
 
\subsection{Effects Caused by a Vector Channel}


Concerning the chiral phase transition, the repulsive vector coupling at very low chemical potential has only little effect over the model's phase diagram with regards to basic thermodynamic quantities such as the pressure, entropy density and energy density.
However, the situation is different if one considers fluctuations which are derivatives of the
pressure. As shown in Ref. \cite{Kunihiro:1991qu}, the presence of a vector coupling suppresses the
quark number susceptibility at high temperature, $T>T_{pc}$. For large chemical potentials, on the
other hand, the repulsive nature of the vector interaction also affects the equation of state and, as a result, it weakens the first order phase transition, lowers the critical temperature while increasing the critical chemical potential. And for a sufficiently strong vector coupling the first order transitions together with the critical point disappear altogether \cite{Fukushima:2008wg}.  

In order to study its effect on the scaled correlation, $\rpp$, let us add the vector channel, $\delta {\cal L} = -G_V(\bar{\psi}\gamma_\mu \psi)^2$, where $G_V$ is the vector coupling constant, to the original Lagrangian (\ref{njl2}),

\begin{equation}
\mathcal{L}_{\rm NJL}={\bar \psi}\left( i{\partial \hbox{$\!\!\!/$}}-m\right) \psi
+G\left[ ({\bar \psi}\psi)^{2}-({\bar{\psi}} \gamma _{5}{\vec{\tau}}\psi
  )^{2}\right] -G_V(\bar{\psi}\gamma_\mu \psi)^2.
\label{njlGv}
\end{equation}

At the mean field level, the thermodynamic potential reads

\beq
\Omega (T,\mu, M,\rho)= \frac{(M-m)^2}{4G} - G_V \rho^2  - 2N_cN_f \int \frac{d^3p}{(2\pi)^3}\left \{E+T\ln \left [1+e^{-(E-\mu+2G_V \rho)/T} \right ] + T\ln \left [1+e^{-(E+\mu-2G_V \rho)/T} \right ] \right \},
\eeq 

\noindent where $\rho$ is the net quark density number. 




To evaluate $\chi_{q,\bar{q}}$ we proceed the same way as in the previous section, i. e. considering $\mu_q$ for particles and $\mu_{\bar q}$ for anti-particles, but noting that the thermodynamical potential is now a function of $\rho$ as well, $\Omega=\Omega (T,\mu_q,\mu_{\bar q},M,\rho)$. Just as before, in order to achieve consistent numerical results, we must set $\mu_q=-\mu_{\bar q}= \mu$. Taking the derivative with respect to $\mu_{\bar q}$ we get 

\beq
\frac{d \Omega}{d \mu_{\bar q}}= \frac{\partial \Omega}{\partial M} \frac{\partial M}{\partial \mu_{\bar q}}+ \frac{\partial \Omega}{\partial \rho} \frac{\partial \rho}{\partial \mu_{\bar q}}+ \frac{\partial \Omega}{\partial \mu_{\bar q}}\, .
\eeq

\noindent Taking the derivative with respect to $\mu_q$ we get

\begin{eqnarray}
\chi_{q,\bar q}&=&-\frac{\partial M}{\partial \mu_{\bar q}}\left ( \frac{\partial^2 \Omega}{\partial M^2}\frac{\partial M}{\partial \mu_q}+\frac{\partial^2 \Omega}{\partial M \partial \rho}\frac{\partial \rho}{\partial \mu_q}+ \frac{\partial^2 \Omega}{\partial M \partial \mu_q} \right )\nonumber\\ 
& &-\frac{\partial \rho}{\partial \mu_{\bar q}}\left ( \frac{\partial^2 \Omega}{\partial \rho^2}\frac{\partial \rho}{\partial \mu_q}+\frac{\partial^2 \Omega}{\partial M \partial \rho}\frac{\partial M}{\partial \mu_q}+ \frac{\partial^2 \Omega}{\partial \rho \partial \mu_q} \right )\nonumber\\
& &-\frac{\partial^2 \Omega}{\partial \mu_{\bar q} \partial M}\frac{\partial M}{\partial \mu_q}-\frac{\partial^2 \Omega}{\partial \mu_{\bar q} \partial \rho}\frac{\partial \rho}{\partial \mu_q}-\frac{\partial^2 \Omega}{\partial \mu_q \partial \mu_{\bar q}}\, .
\vspace{0,4 cm}
\label{EqChiGv}
\end{eqnarray}

Most of the terms in the above equation may be simplified. This is done in Appendix \ref{AppXqqVector}, where we obtain an expression for $\chi_{q,\bar{q}}$ that can be numerically evaluated. This allows us to understand how the proton--anti-proton scaled correlation depends on the strength of the vector coupling, $G_V$. 

\begin{figure}[htb]
\begin{center}
\includegraphics[height=5.3cm]{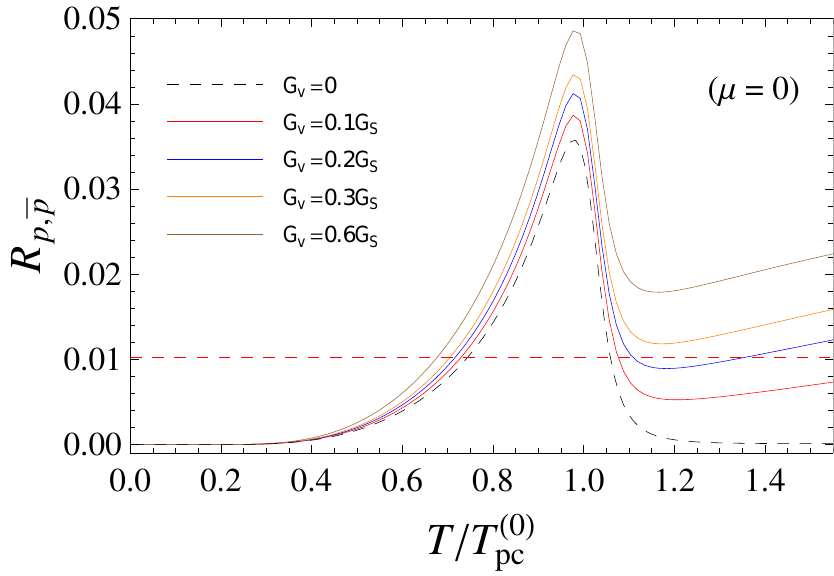}
\includegraphics[height=5.3cm]{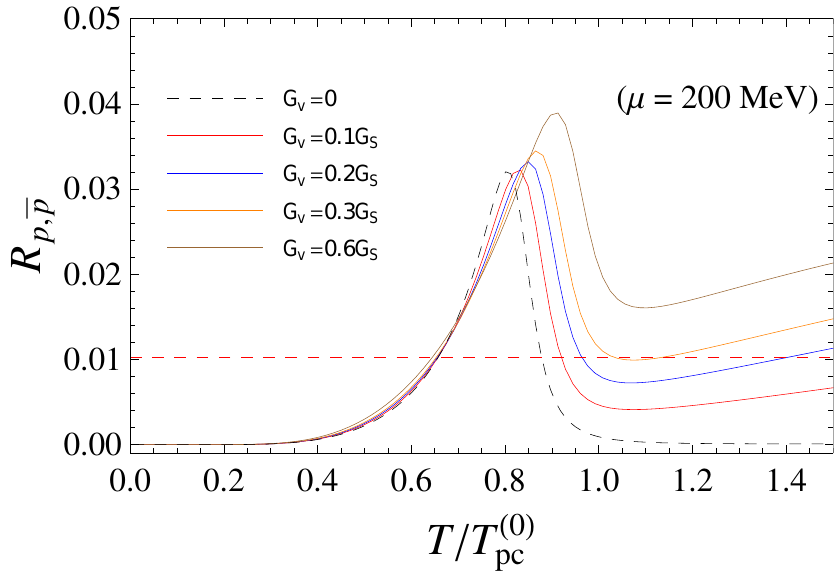}
\caption{Proton--anti-proton scaled correlation as a function of temperature normalized by the pseudo-critical temperature at zero chemical potential. Left panel: The $\mu=0$ case. Right panel: The $\mu=200\, \rm MeV$ case. The horizontal dashed line represents a linear fit to the  STAR data from Fig.~\ref{Fig_RppData}.}
\label{Fig_Gv_RvsT_Mu0}
\end{center}
\end{figure}

In Fig. \ref{Fig_Gv_RvsT_Mu0} we show the scaled correlation,  $R_{p,\bar{p}}$, as a function of $T/T_{pc}^{(0)}$
for $\mu=0$ (left panel) and $G_{V}$ values ranging from  $G_V/G=0.1,\ldots 0.3$ and a higher value, $G_V/G=0.6$, that is particularly different from the other values because its corresponding phase diagram shows no first order transition and no critical point\footnote{This feature actually depends on the choice of the model parameters. Other parametrization may lead to different phase diagrams for the same relations $G_V/G$ used in this work.}. This choice covers the
value of $G_V/G = 0.33$ which was determined in Ref. \cite {Sugano:2014pxa} using LQCD constraints. For comparison, the (dashed) line corresponding to the  $G_V=0$ case is also presented. 

We can clearly see that the peak near the pseudo-critical temperature ($T_{pc}^{(0)} \simeq 188\, \rm MeV$ for all lines) increases as the vector coupling increases. 
For temperatures $T$ above the pseudo-critical temperature,  $T>T_{pc}^{(0)}$, we observe  a curious behavior:  for vanishing vector coupling $\rpp$ drops rapidly and then slowly decreases  towards  zero  (dashed line), but for finite vector coupling, $G_V > 0$, on the other hand, after a sharp drop, $\rpp$ steadily increases with $T$. This means that the scaled correlation is influenced by the vector coupling at high $T$ even at zero chemical potential. 
This can be understood by noting that for the net quark number susceptibility the vector interaction
is repulsive, and thus it gets screened. On the other hand, for quark--anti-quark pairs, the vector
interaction is attractive, and thus the correlation is enhanced. Even at $\mu=0$ we have quarks and
anti-quarks and due to the vector interaction they get more correlated\footnote{We have verified by
  explicit calculation that indeed the net-quark susceptibility is screened as discussed in
  \cite{Kunihiro:1991qu}. We furthermore have checked that the relation between the various
  variances, Eq.~\eqref{eq:k2_prot}, holds in the model calculation. }. The scalar interaction, on the other hand, decreases with decreasing effective quark mass, and, thus, gets weaker for temperatures above $T_c$. We note that in the limit $G_V \rightarrow 0$ we recover the results obtained in the previous section, i. e., the dashed black line in Fig. \ref{Fig_Gv_RvsT_Mu0}.

\begin{figure}[htb]
\begin{center}
\includegraphics[height=5.3cm]{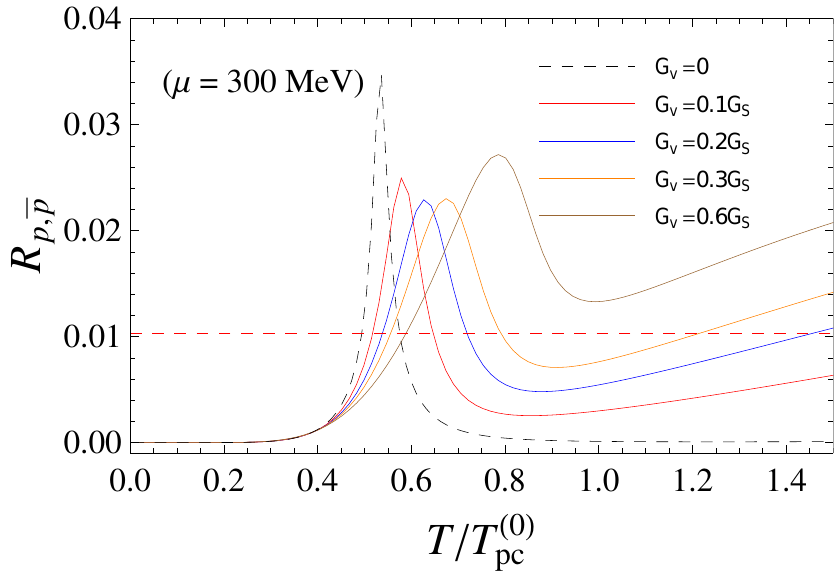}
\includegraphics[height=5.3cm]{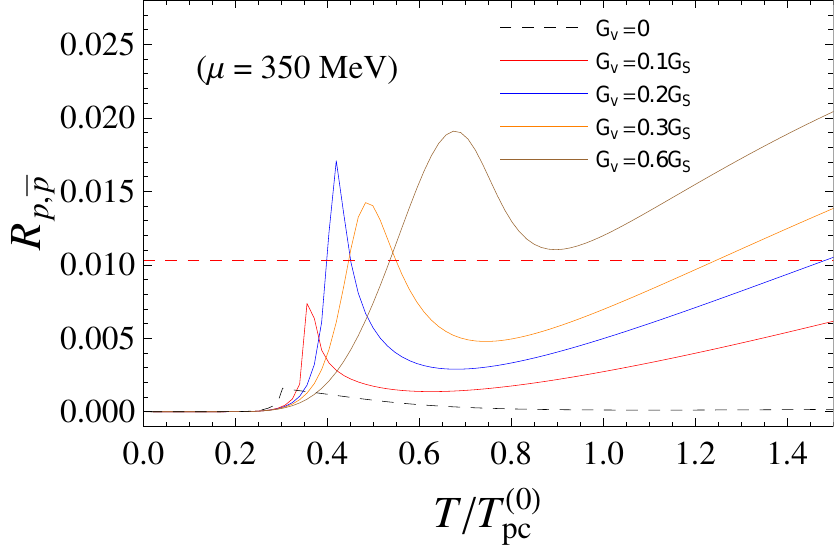}
\caption{Proton--anti-proton scaled correlation as a function of $T/T_{pc}^{(0)}$ at different values of $G_V$. Left panel: $\mu=300 \mev$. Right panel: $\mu=350\, \rm MeV$. The horizontal dashed line represents a linear fit from STAR data, Fig. \ref{Fig_RppData}.}
\label{Fig_Gv_RvsT_Mu300}
\end{center}
\end{figure}

With increasing  chemical potential, $\mu$, the vector coupling gives rise to additional changes in $R_{p,\bar{p}}$. This is shown in the right panel of Fig. \ref{Fig_Gv_RvsT_Mu0}, where we consider $\mu= 200 \mev$, and in Fig.~\ref{Fig_Gv_RvsT_Mu300} where we show the results for  $\mu=300 \mev$ and $\mu = 350 \mev$. We find that with increasing vector coupling the peaks of $\rpp$ shift towards higher temperature. In addition the peaks become less sharp and closer to the linear fit of the STAR data, represented by the horizontal dashed line. Both effects are stronger at higher values of $\mu$. In addition, while for $\mu = 0$ and $\mu=200 \,\rm MeV$ the peaks at finite $G_{V}$ are larger than that for $G_{V}=0$, we observe the opposite for $\mu=300 \mev$. The reason for this behavior is that  $G_V$ shifts the critical chemical potential to higher values. Consequently, for $G_{V}>0$ and $\mu=300\mev$ we are still sufficiently far away from the critical region whereas for $G_{V}=0$ we are close and thus see the critical enhancement of $\rpp$. 
At $\mu=350\, \rm MeV$ (right panel of Fig. \ref{Fig_Gv_RvsT_Mu300}) we see that the cases
$G_V=0.1\, G$ and $G_V=0.2\, G$ also show sharp peaks due to the fact that these values for the
chemical potential are close to  the respective critical values. The result for  $G_V=0$  (dashed line), on the other hand, does not exhibit a significant peak since in this case we are beyond the critical chemical potential. In general, we see that the peaks of $R_{p,\bar{p}}$ are lower at $\mu=350\, \rm MeV$ than in the previous cases ($\mu < 350 \, \rm MeV$). %

\begin{figure}[htb]
\begin{center}
\includegraphics[height=5.5cm]{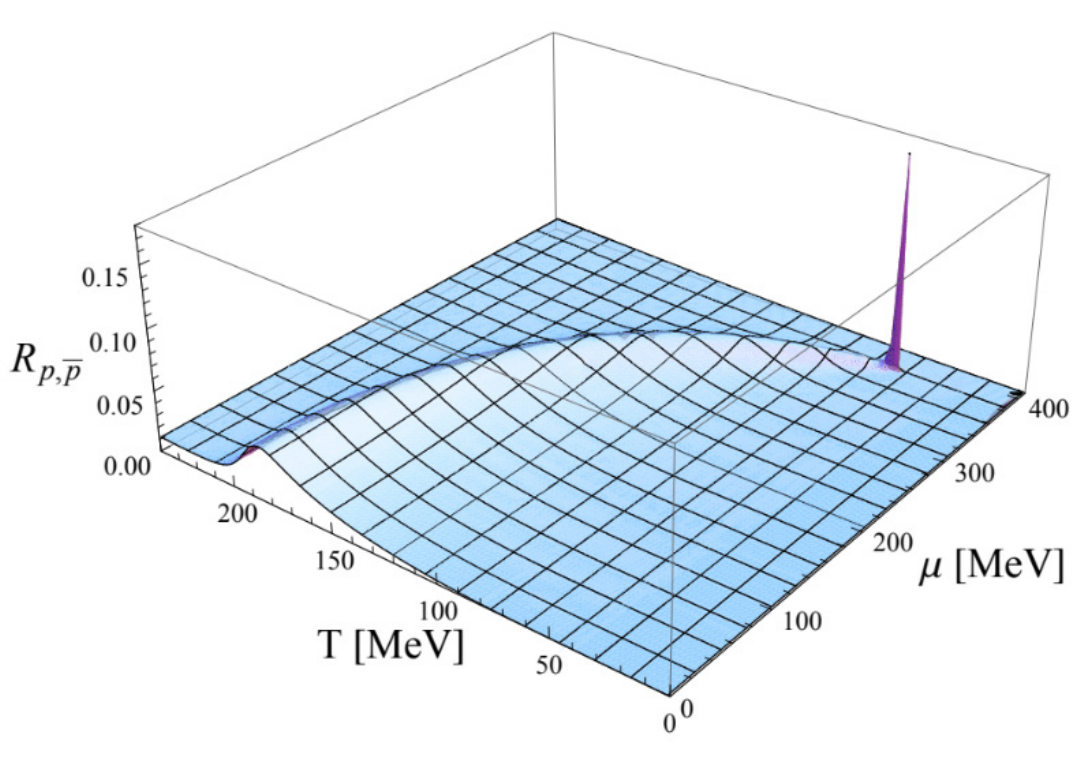}
\includegraphics[height=5.5cm]{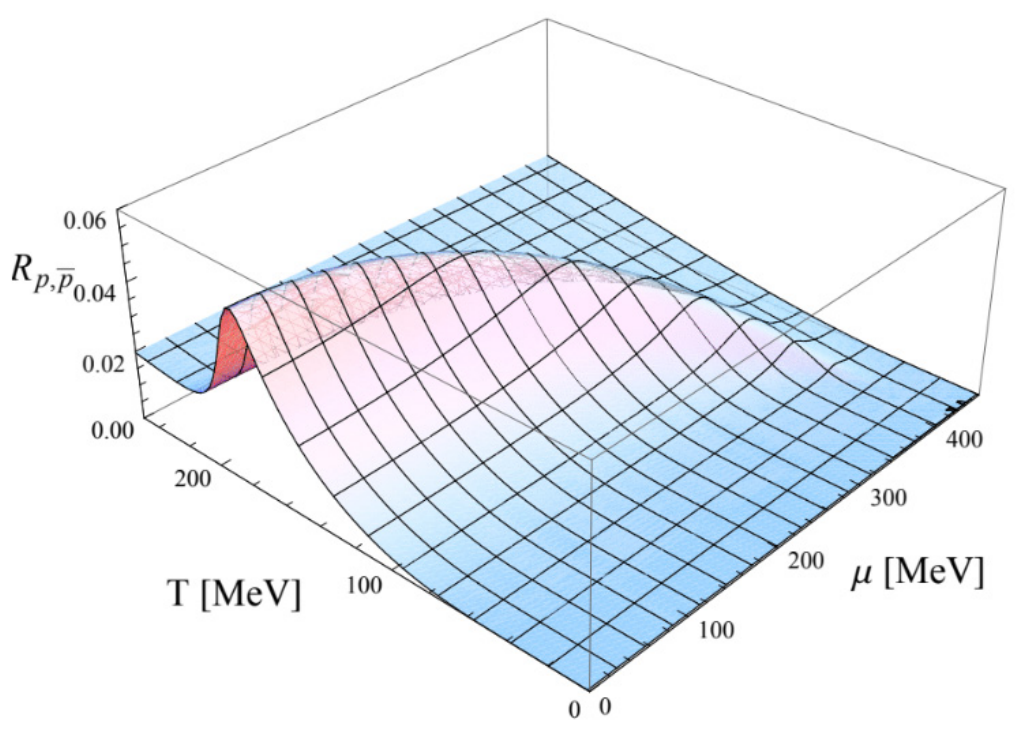}
\caption{Proton--anti-proton scaled correlation in the $T-\mu$ plane for $G_V=0.3\, G$ (left), where we can see a peak at the critical point, and $G_V=0.6\, G$ (right), where there is no critical point.}
\label{Fig_Gv03_R3D}
\end{center}
\end{figure}

Extending the analysis to the whole $T-\mu$ plane we obtain a better overall view of the vector
coupling effects on $R_{p,\bar{p}}$. Fig. \ref{Fig_Gv03_R3D} (left panel) shows the
proton--anti-proton scaled correlation in the $T-\mu$ plane for $G_V=0.3\, G$. We  observe a general
similarity with Fig. \ref{Fig_Gv0_R3D}. If we increase the vector coupling even more the first order phase transition turns into a crossover, meaning that there is no critical point in the phase diagram and, therefore, no
peak in the correlation, as the right panel of Fig. \ref{Fig_Gv03_R3D} shows for $G_V=0.6\,
G$.

\begin{figure}[htb]
\begin{center}
\includegraphics[height=5.5cm]{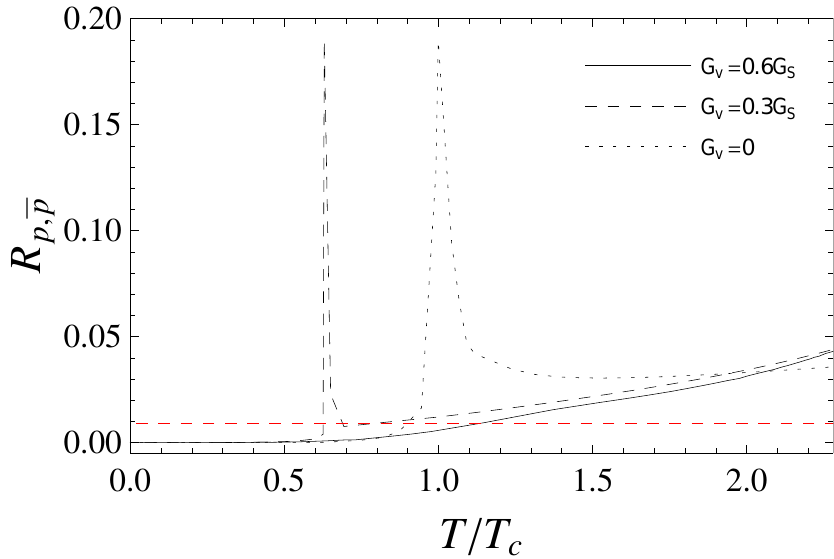}
\caption{The proton--anti-proton scaled correlation along the (pseudo)-critical line for different values of $G_V$. The temperature is normalized by the critical temperature, $T_c$, for $G_V=0$. The horizontal dashed line in red represents a linear fit from STAR data.}
\label{Fig_Gv_RvsT_PD}
\end{center}
\end{figure}

Finally, in order to compare the behavior of $R_{p,\bar{p}}$ along the phase transition border, we present
Fig. \ref{Fig_Gv_RvsT_PD} which shows $R_{p,\bar{p}}$ along the (pseudo)-critical line for $G_V=0$, $G_V=0.3\, G$ and $G_V=0.6\, G$. Contrary to the case of vanishing vector coupling, all the results for finite $G_{V}$ show a sizable increase of $\rpp$ with temperature along the pseudo-critical line. This is certainly not seen in the STAR data (see Fig.~\ref{Fig_RppData}), although the error-bars may be still to large to make a definitive statement at this time. However, the effect of the vector coupling is again rather generic. Thus, an improved  measurement of $\rpp$ could be able to put a limit on the strength of the vector coupling at the freeze out conditions.
We see that the peaks at the critical point look very similar but we should notice that the height of the peak actually depends on the numerical resolution used in the evaluations. 
 




\section{Summary and Conclusions}
\label{conclude}

Inspired by the STAR data on cumulants of proton and anti-proton distributions, we evaluated the
quark--anti-quark scaled correlation within the NJL model for light quarks in the mean field
approximation, which allowed us to estimate the proton--anti-proton scaled correlation. At low
chemical potential, which is equivalent to high energy collision, we found that the correlation
increases with $T$ and has a smooth peak at the pseudo-critical temperature. Extending the
calculation to finite values of the baryon number chemical potential, we found that the scaled
correlation function always exhibits a maximum along the pseudo-critical line. The value of this maximum is
rather constant before it diverges close to the critical point. The scaled correlation extracted
from the STAR data is, within errors, also rather constant suggesting that the freeze out happens
close to the pseudo-critical line. However, the calculated value for the scaled correlation is about
a factor of four larger than the STAR data, which may very well be due to the limits of the effective
model employed here.

We further studied the effect of a repulsive vector channel which is known to disfavor first order
phase transitions. At low chemical potentials we saw that the inclusion of the vector interaction
enhanced the correlation. This is understandable, since the vector interaction is attractive
for particle--anti-particle pairs. At the critical point we saw a peak similar to the case with no vector interaction but located at a lower critical temperature and a higher critical chemical potential, as expected. However, in the case that the repulsive vector coupling was strong enough to suppress the first order transition, we found no critical point in the phase diagram and, therefore, no peak in the correlation.  

 
Although the present calculation was carried out in the mean field approximation, we believe that
the qualitative observations are robust: The presence of critical point should result in a peak in
the scaled correlation, $R_{p,\bar{p}}$ and thus direct a measurement of $R_{p,\bar{p}}$ rather 
than indirect extraction performed here, would be very valuable. In particular, at the lowest
energies of the RHIC beam energy scan, $\sqrt{s}=7.7\gev$, such a measurement would be very
welcome. At this energy,  preliminary STAR data with
increased acceptance, exhibit a strong increase in the fourth order net-proton number cumulant
\cite{Luo:2015ewa} which are due to large four-proton correlations \cite{Bzdak:2016sxg}. These 
correlations are not easily explained by conventional scenarios
\cite{Bzdak:2016jxo,Bzdak:2018uhv}. Unfortunately, the  present extraction of $R_{p,\bar{p}}$ from the difference of net-proton and proton and anti-proton variance at present has too large an error especially at the low energies to allow for any conclusion.

Finally we note that, although $R_{p,\bar{p}}$ is in general not well defined in thermal field
theory, it is a bonafide observable, since it is obviously accessible in experiment. Thus it might be worthwhile to develop appropriate projections of a thermal system onto asymptotic states in order to access this observable also in model/calculations which go beyond the mean field approximation such as Dyson-Schwinger or functional re-normalization (FRG) group methods \cite{Fischer:2009gk,Herbst:2010rf}.

\acknowledgments

A.F.G. and M.B.P. are grateful to Coordena\c c\~{a}o de Aperfei\c coamento de Pessoal de N\'{\i}vel Superior
(CAPES-Brazil) and Conselho Nacional de Desenvolvimento Cient\'{\i}fico e Tecnol\'{o}gico
(CNPq-Brazil) for financial support. V.K. was supported by the U.S. Department of Energy, Office of
Science, Office of Nuclear Physics, under contract number DE-AC02-05CH11231. This work is  a part of the project CNPq-INCT-FNA Proc. No. 464898/2014-5. This work also received support within the framework of the Beam Energy Scan Theory (BEST) Topical Collaboration.

\appendix

\section{Evaluation of $\chi_{q,\bar q}$}
\label{AppXq}

In this appendix we show, in a more detailed fashion, how to  obtain the expression given by Eq. (\ref{Chi2}). One may start by considering the first order derivative

\beq
\frac{d \Omega}{d \mu_{\bar{q}}}=\frac{\partial \Omega}{\partial \mu_{\bar{q}}}+\frac{\partial \Omega}{\partial M}\frac{\partial M}{\partial \mu_{\bar{q}}}\, ,
\eeq
which allows for the off-diagonal susceptibility to be obtained once  an additional derivative with
respect to $\mu_q$ is taken 

\beq
\chi_{q,\bar{q}}=-\frac{\partial^2 \Omega}{\partial \mu_q \partial \mu_{\bar{q}}}-\frac{\partial^2 \Omega}{\partial M^2}\frac{\partial M}{\partial \mu_q}\frac{\partial M}{\partial \mu_{\bar{q}}}-\frac{\partial^2 \Omega}{\partial M \partial \mu_q}\frac{\partial M}{\partial \mu_{\bar{q}}}-\frac{\partial^2 \Omega}{\partial M \partial \mu_{\bar{q}}}\frac{\partial M}{\partial \mu_q}\, .
\vspace{0,4 cm}
\label{Eq_Chi}
\eeq

Next we work out each term in the above expression. Recalling that $\mu_q$ and $\mu_{\bar{q}}$ contribute to distinct terms in Eq. (\ref {Omega2}) one sees that the first term on the r.h.s. of Eq.(\ref {Eq_Chi}) vanishes. In the second term we identify

\beq
\frac{\partial^2 \Omega}{\partial M^2}=\Omega^{\prime \prime}_M
\vspace{0,4 cm}
\label{Eq_Omega2M}
\eeq

\noindent with the curvature of the thermodynamic potential evaluated at the gap solution. The term $\partial M/ \partial \mu_q$ is non-trivial. However, we can make use of the gap equation,

\beq
\frac{\partial \Omega}{\partial M}=0\, ,
\vspace{0,4 cm}
\eeq 

\noindent to write

\beq
\frac{d}{d\mu_q}\left ( \frac{\partial \Omega}{\partial M} \right) = \frac{\partial^2 \Omega}{\partial M^2}\frac{\partial M}{\partial \mu_q}+\frac{\partial^2 \Omega}{\partial \mu_q \partial M}=0 \, .
\vspace{0,4 cm}
\eeq

\noindent Isolating $\partial M/ \partial \mu_q$ and using Eq. (\ref{Eq_Omega2M}) we get

\beq
\frac{\partial M}{\partial \mu_q}=-\frac{1}{\Omega^{\prime \prime}_M}\frac{\partial^2 \Omega}{\partial M \partial \mu_q}\, ,
\vspace{0,4 cm}
\label{Eq.DMDMU}
\eeq

\noindent noting that the same relation is valid upon replacing $q\rightarrow \bar{q}$. Then, substituting Eq. (\ref{Eq.DMDMU}) into Eq. (\ref{Eq_Chi}) yields

\beq
\chi_{q, \bar{q}}=\frac{1}{\Omega^{\prime \prime}_M}\frac{\partial^2 \Omega}{\partial M \partial \mu_q}\frac{\partial^2 \Omega}{\partial M \partial \mu_{\bar{q}}}\, .
\vspace{0,4 cm}
\label{Eq_Chi2}
\eeq

We can simplify the above equation further by remembering that (see Eq. (\ref {Eq_Densities} ))

\beq
\frac{\partial \Omega}{\partial \mu_q}=-\rho_q \,\,\,\,\,\,\,\,\,\,\, \mbox{and} \,\,\,\,\,\,\,\,\,\,\, \frac{\partial \Omega}{\partial \mu_{\bar{q}}}=-\rho_{\bar{q}}\, ,
\vspace{0,4 cm}
\eeq

\noindent so that Eq. (\ref{Eq_Chi2}) turns into 

\beq
\chi_{q, \bar q}=\frac{1}{\Omega^{\prime \prime}_M}\frac{\partial \rho_q}{\partial M}\frac{\partial \rho_{\bar q}}{\partial M}\, ,
\eeq

\noindent which has been used to write Eq. (\ref{Eq_Chi3}).


\section{Evaluation of $\chi_{q, \bar{q}}$ with Vector Channel}
\label{AppXqqVector}
Let us now put the expression for $\chi_{q, \bar{q}}$, given by Eq. (\ref{EqChiGv}), into a form which will suit  the implementation of numerical routines. 
We can start by simplifying the $\partial M/ \partial \mu_q$ term recalling that, as usual, the gap equation is simply

\beq
\frac{\partial \Omega}{\partial M}=0 \, .
\vspace{0,4 cm}
\eeq

\noindent Taking the derivative with respect to $\mu_q$ and remembering that now $\Omega$ is also a function of $\rho$  we get

\beq
\frac{d}{d \mu_q}\left ( \frac{\partial \Omega}{\partial M} \right ) = \frac{\partial^2 \Omega}{\partial M^2} \frac{\partial M}{\partial \mu_q}+ \frac{\partial^2 \Omega}{\partial M \partial \rho} \frac{\partial \rho}{\partial \mu_q}+ \frac{\partial^2 \Omega}{\partial M \partial \mu_q}=0 \, .
\vspace{0,4 cm}
\label{Eq.I}
\eeq


Proceeding in the same way for the $\partial \rho / \partial \mu_q$ term and upon using the stationary condition

\beq
\frac{\partial \Omega}{\partial \rho} =0\, ,
\vspace{0,4 cm}
\eeq

\noindent allow us to write 

\beq
\frac{d}{d \mu_q}\left ( \frac{\partial \Omega}{\partial \rho} \right ) = \frac{\partial^2 \Omega}{\partial \rho^2}\frac{\partial \rho}{\partial \mu_q}+ \frac{\partial^2 \Omega}{\partial \rho \partial M}\frac{\partial M}{\partial \mu_q} + \frac{\partial^2 \Omega}{\partial \rho\partial \mu_q}=0\, ,
\vspace{0,4 cm}
\label{Eq.III}
\eeq



\noindent At this stage we introduce the following notation for the second order derivatives

\beq
\frac{\partial^2 \Omega}{\partial M^2}=\Omega^{\prime \prime}_M \,, \,\,\,\,\,\,\,\,\, \frac{\partial^2 \Omega}{\partial \rho^2}=\Omega^{\prime \prime}_\rho \,, \,\,\,\,\,\,\,\,\, \frac{\partial^2 \Omega}{\partial M \partial \rho}= \Omega^{\prime \prime}_{M\rho}\, ,
\vspace{0,4 cm}
\eeq

\noindent so that Eq. (\ref{Eq.I}) may be written as

\beq
\frac{\partial^2 \Omega}{\partial \mu_q \partial M}+ \Omega^{\prime \prime}_M \frac{\partial M}{\partial \mu_q}+\Omega^{\prime \prime}_{M\rho}\frac{\partial \rho}{\partial \mu_q}=0\, ,
\vspace{0,4 cm}
\eeq



\noindent while Eq. (\ref{Eq.III}) reads

\beq
\frac{\partial^2 \Omega}{\partial \mu_q \partial \rho}+\Omega^{\prime \prime}_{\rho}\frac{\partial \rho}{\partial \mu_q}+\Omega^{\prime \prime}_{M\rho}\frac{\partial M}{\partial \mu_q}=0\, .
\vspace{0,4 cm}
\eeq



\noindent These manipulations allow us to obtain $\partial \rho / \partial \mu_q$ from

\beq
\frac{\partial \rho}{\partial \mu_q}=-\frac{1}{\Omega^{\prime \prime}_{\rho}}\left [ \frac{\partial^2 \Omega}{\partial \mu_q \partial \rho}+\Omega^{\prime \prime}_{M\rho}\frac{\partial M}{\partial \mu_q} \right ]\, .
\vspace{0,4 cm}
\label{Eq.III2}
\eeq

\noindent Note that the same is valid for $\mu_{\bar q}$ (switching $\mu_q$ for $\mu_{\bar q}$ in the above equation). Next, substituting Eq. (\ref{Eq.III2}) into Eq. (\ref{Eq.I}) we get

\beq
\frac{\partial^2 \Omega}{\partial \mu_q \partial M}+\Omega^{\prime \prime}_M \frac{\partial M}{\partial \mu_q}+\Omega^{\prime \prime}_{M\rho} \left[-\frac{1}{\Omega^{\prime \prime}_{\rho}}\left (\frac{\partial^2 \Omega}{\partial \mu_q \partial \rho}+\Omega^{\prime \prime}_{M\rho}\frac{\partial M}{\partial \mu_q} \right) \right]=0\, .
\vspace{0,4 cm}
\eeq

\noindent Isolating $\partial M / \partial \mu_q$ in the above expression we find



\beq
\frac{\partial M}{\partial \mu_q}=\left[ \Omega^{\prime \prime}_{M}-\frac{(\Omega^{\prime \prime}_{M\rho})^2}{\Omega^{\prime \prime}_{\rho}}\right]^{-1}\left[ \frac{\Omega^{\prime \prime}_{M\rho}}{\Omega^{\prime \prime}_{\rho}}\frac{\partial^2 \Omega}{\partial \mu_q \partial \rho}-\frac{\partial^2 \Omega}{\partial \mu_q \partial M} \right]\, ,
\vspace{0,4 cm}
\label{Eq.V}
\eeq



\noindent which can be substituted in Eq. (\ref{Eq.III}) to produce  an expression for $\partial \rho/\partial \mu_q$ that can be numerically evaluated. Namely,

\beq
\frac{\partial \rho}{\partial \mu_q}=-\frac{1}{\Omega^{\prime \prime}_{\rho}}\left[ \frac{\partial^2 \Omega}{\partial \rho \partial \mu_q}+\Omega^{\prime \prime}_{M\rho} \left(\frac{\Omega^{\prime \prime}_{M\rho}}{\Omega^{\prime \prime}_{\rho}}\frac{\partial^2 \Omega}{\partial \rho \partial \mu_q}-\frac{\partial^2 \Omega}{\partial M \partial \mu_q}\right)\left( \Omega^{\prime \prime}_M -\frac{(\Omega^{\prime \prime}_{M\rho})^2}{\Omega^{\prime \prime}_{\rho}}\right)^{-1}\right]\, .
\vspace{0,4 cm}
\label{Eq.VII}
\eeq
Now, we may introduce a new variable to further simplify the notation 

\beq
n_q=\frac{\partial \Omega}{\partial \mu_q} \,  ,
\vspace{0,4 cm}
\eeq

\noindent so that Eq. (\ref{Eq.VII}) becomes

\beq
\rho^{\prime}_q=\frac{\partial \rho}{\partial \mu_q}= - \frac{1}{\Omega^{\prime \prime}_{\rho}} \left [ \frac{\partial n_q}{\partial \rho}+ \Omega^{\prime \prime}_{M \rho} \left ( \frac{\Omega^{\prime \prime}_{M\rho}}{\Omega^{\prime \prime}_{\rho}}\frac{\partial n_q}{\partial \rho}-\frac{\partial n_q}{\partial M}\right ) \left ( \Omega^{\prime \prime}_M - \frac{(\Omega^{\prime \prime}_{M \rho})^2}{\Omega^{\prime \prime}_{\rho}} \right )^{-1} \right ] \, , 
\vspace{0,4 cm}
\eeq

\noindent while Eq. (\ref{Eq.V}) can be written as

\beq
M^{\prime}_q=\frac{\partial M}{\partial \mu_q}=\left[ \Omega^{\prime \prime}_{M}-\frac{(\Omega^{\prime \prime}_{M\rho})^2}{\Omega^{\prime \prime}_{\rho}}\right]^{-1}\left[ \frac{\Omega^{\prime \prime}_{M\rho}}{\Omega^{\prime \prime}_{\rho}}\frac{\partial n_q}{\partial \rho}-\frac{\partial n_q}{\partial M} \right]\, .
\vspace{0,4 cm}
\eeq

\noindent In the case of anti-particles, we find the same results upon replacing $q \rightarrow \bar q$. Finally, the off diagonal susceptibility in the presence of a vector coupling may be written as 

\beq
\chi_{q,\bar q}= M^{\prime}_{\bar q} \left ( \Omega^{\prime \prime}_M M^{\prime}_q + \Omega^{\prime \prime}_{M\rho}\rho^{\prime}_q + \frac{\partial n_q}{\partial M} \right ) + \rho^{\prime}_{\bar q} \left ( \Omega^{\prime \prime}_{\rho}\rho^{\prime}_q + \Omega^{\prime \prime}_{M\rho}M^{\prime}_q + \frac{\partial n_q}{\partial \rho} \right ) + \frac{\partial n_{\bar q}}{\partial M} M^{\prime}_q + \frac{\partial n_{\bar q}}{\partial \rho}\rho^{\prime}_q + \frac{\partial^2 \Omega}{\partial \mu_q \partial \mu_{\bar q}}\, .
\vspace{0,4 cm}
\label{EqChiGvFinal}
\eeq 
which allows us to numerically evaluate the particle--anti-particle scaled correlation, given by Eq. (\ref{Eq_R2}), in a conveniently  way.

\bibliography{V8_arXiv}

\end{document}